\documentclass[10pt]{article} 

\usepackage[accepted]{tmlr}
\usepackage{hyperref}
\usepackage{url}
\usepackage{natbib}
\usepackage{tikz}
\usepackage{xcolor}
\definecolor{graphicbackground}{rgb}{0.96,0.96,0.8}
\definecolor{codebackground}{rgb}{0.9,0.9,1}
\usepackage{graphicx}
\usepackage{comment}

\usepackage{geometry}
\usepackage{changepage}

\title{Unlocking the conversion of Web Screenshots into HTML Code with the WebSight Dataset}

\begin{document}

\maketitle

\vspace{-15mm}
\textbf{\sffamily{Hugo Laurençon, Léo Tronchon, Victor Sanh}} \\
[1ex]Hugging Face

\vspace{3mm}
\begin{adjustwidth}{0.5cm}{0.5cm}
Using vision-language models (VLMs) in web development presents a promising strategy to increase efficiency and unblock no-code solutions: by providing a screenshot or a sketch of a UI, a VLM could generate the code to reproduce it, for instance in a language like HTML. Despite the advancements in VLMs for various tasks, the specific challenge of converting a screenshot into a corresponding HTML has been minimally explored. We posit that this is mainly due to the absence of a suitable, high-quality dataset. This work introduces WebSight, a synthetic dataset consisting of 2 million pairs of HTML codes and their corresponding screenshots. We fine-tune a foundational VLM on our dataset and show proficiency in converting webpage screenshots to functional HTML code. To accelerate the research in this area, we open-source WebSight.
\end{adjustwidth}

\vspace{1mm}

\sffamily{
Dataset: \url{https://huggingface.co/datasets/HuggingFaceM4/WebSight}\\
}

\normalfont

\begin{figure}[h]
    \centering
    \includegraphics[width=1.0\textwidth]{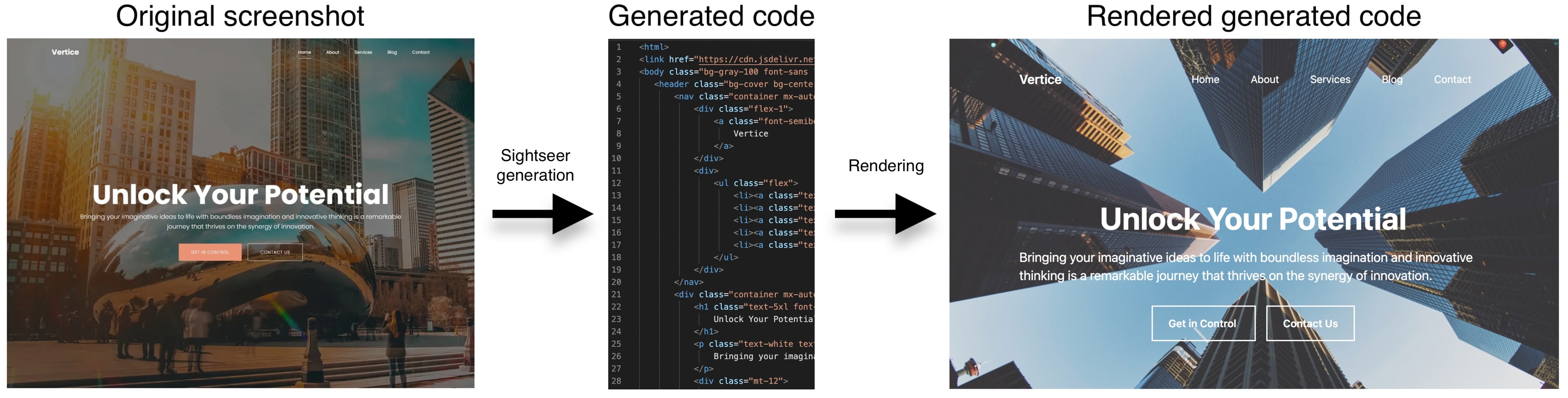}
    \caption{Comparison of an original web page (input) on the left, and the rendering of the code generated by our model - Sightseer - (output) on the right. To match the visual aspect of the original webpage, the model picked a suitable image background from \url{www.unsplash.com}}
    \label{fig:main_generations}
\end{figure}

\section{Introduction}

Current advancements in vision-language models (VLMs) have significantly improved their capabilities, enabling them to master a variety of tasks including image captioning, question answering, and optical character recognition (OCR) \citep{openai2023gpt4, geminiteam2023gemini, hong2023cogagent, liu2024llavanext}. Despite these achievements, the task of converting screenshots of websites or web components into usable HTML code—a process highly valuable to web developers—remains relatively unexplored, particularly in the open-source community. The development and open-source release of a model capable of such a conversion could unlock new AI-powered tools for UI developers, facilitating the creation of no-code modules and plugins for design tools like Figma. For instance, the ability to rapidly transform a design sketch into a functional UI component and code could significantly increase the iteration pace for UI developers.

We posit that the primary challenge for VLMs to achieve proficiency in this specific task does not stem from the inherent difficulty of the task itself. Rather, it is the lack of a large, high-quality, dataset of pairs of HTML codes and their associated screenshots that poses the primary obstacle. In fact, VLMs are commonly trained on web-scale datasets of image-text pairs \citep{schuhmann2022laion5b, gadre2023datacomp} or multimodal web documents \citep{laurencon2023obelics, zhu2023multimodal}. Having such a dataset of screenshots-HTML pairs as an open and accessible artifact would significantly accelerate research in this area by enabling the community to inspect the data, its limitations and improve upon the dataset. Consequently, our initial focus is on developing a dataset useful for the fine-tuning of VLMs for this task. To accomplish this, several strategies can be considered:

\begin{enumerate}
\item \textit{Leveraging existing webpages and their HTML codes.} The vast repository of HTML files available online (and often captured in web crawls like Common Crawl) presents a tempting resource for generating pairs of screenshots and corresponding HTML codes by simply rendering the HTML and capturing the output. However, this approach poses significant challenges. HTML files found on the web are often laden with noise like comments, scripts or data, and can be excessively lengthy, encompassing a very large number tokens, sometimes even exceeding the maximum sequence length of most current models. This complexity hinders a model's ability to accurately learn the correlation between the contents of a screenshot and the underlying HTML syntax. Additionally, HTML codes frequently incorporate references to external JavaScript (JS) or Cascading Style Sheets (CSS) scripts, or rely on files located in separate directories. This dependency complexifies the creation of a self-contained HTML file that faithfully reproduces the intended design in a screenshot. Given these obstacles, we opted to forego this method in favor of a more controlled approach.

\item \textit{Synthesizing HTML codes with Large Language Models (LLMs).} The most recentlarge language models, especially those trained extensively on programming languages, show remarkable proficiency in generating high-quality code applicable to various domains, including website development. This capability opens the door to artificially create a vast corpus of HTML codes using a LLM specialized in coding which has been further fine-tuned to follow instructions. By adapting the prompts, we can introduce specific constraints to the code generation process, such as controlling the topic, the text length or the image placement in the websites. This level of control not only ensures the production of relevant HTML code but also makes them more suitable for VLMs by providing the models with cleaner, more concise, and structured data that models can be effectively trained on. Our study adopts this approach.
\end{enumerate}

In response to the identified gap, we develop WebSight, a comprehensive synthetic dataset comprising 2 million examples of HTML code paired with corresponding screenshots. Leveraging this dataset, we proceed to fine-tune our forthcoming foundational VLM of 8 billion parameters, notably enhanced by robust OCR capabilities, to obtain the specialized model Sightseer. This fine-tuning process yields promising outcomes, demonstrating the model's proficiency in converting webpage screenshots into functional HTML code. Remarkably, the model also exhibits the versatility to adapt to untrained scenarios, such as transforming handwritten sketches into functional HTML code. To accelerate advancements in this direction, we open source WebSight.

\section{Related work}

\citet{Nguyen2015ReverseEM} uses a classical pipeline of interface elements recognition (images, texts, containers, etc.) with computer vision and optical character, followed by heuristics to generate code on these detections. The authors show the effectiveness of this approach on mobile UIs. \citet{beltramelli2017pix2code} introduces an end-to-end method for generating computer code from graphical user interface (GUI) screenshots using deep learning. The model, trained end-to-end, can generate code for different platforms (iOS, Android, and web) from a single input image. It uses convolutional and recurrent neural networks to interpret GUI screenshots and generate corresponding code. In \citet{pix2struct}, authors pre-train a VLM to convert masked screenshots of web pages into simplified HTML, and show the effectiveness of this training objective to pretrain foundational VLM that transfers well to a variety of downstream tasks. Similar to Sightseer, their model accepts images of varying resolutions as input.

In our recent beta release of WebSight-v0.1, we provided a dataset with 823K synthetic pairs of screenshots and associated HTML + traditional CSS code. In the current version of WebSight discussed in this paper (v0.2), we introduce significant improvements. First, WebSight-v0.2 replaces the colored rectangles used as image placeholders in WebSight-v0.1 with real images that match the website's content. Additionally, we adopt Tailwind CSS to streamline the code and facilitate the creation of visually appealing designs. Other notable upgrades include 2.5x the dataset size, offering higher resolution screenshots, and providing richer metadata.

WebSight-v0.1 has already proven to be a helpful resource. In Design2Code \citep{si2024design2code}, the authors create a benchmark for evaluating VLMs at generating HTML code given a screenshot. They also fine-tune an 18B-parameter VLM on WebSight-v0.1, after observing that models trained on synthetic examples outperform those trained on longer, more complex real-world code data.

\section{Construction of the dataset}

\begin{figure}[ht]
\centering
\includegraphics[width=1.0\textwidth]{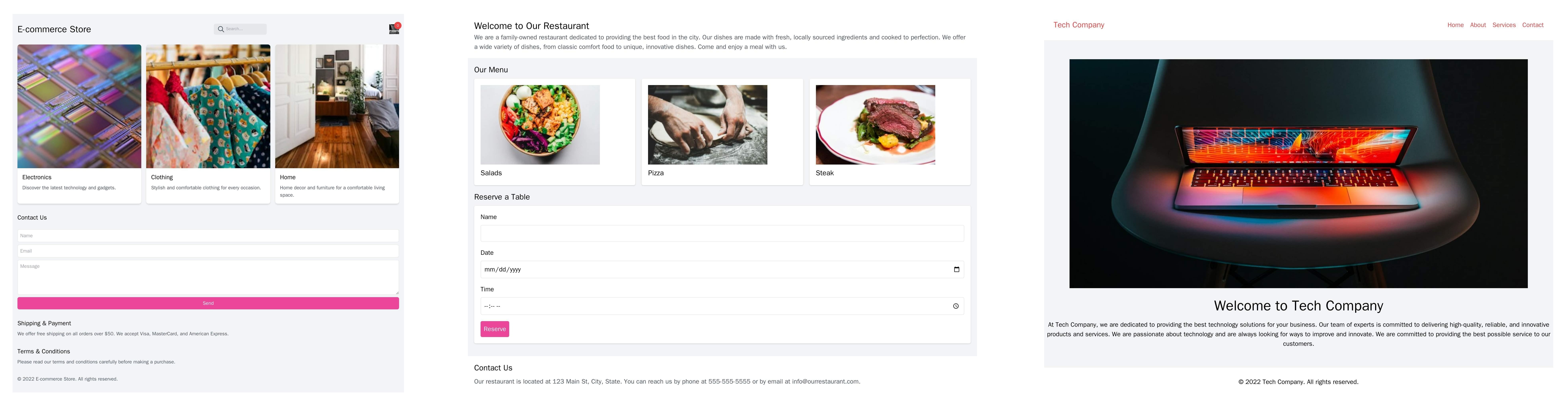}
\includegraphics[width=0.55\textwidth]{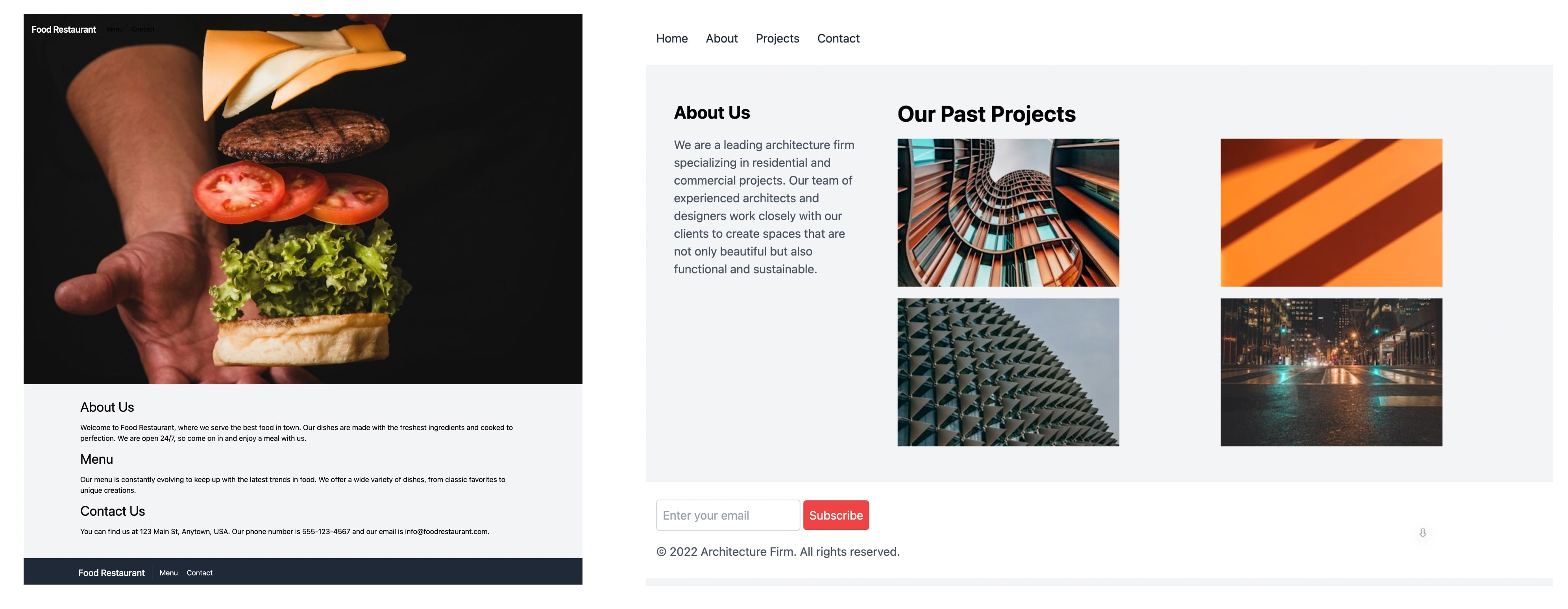}
\caption{Examples of synthetic web pages present in WebSight.}
\label{fig:ex_wesight}
\end{figure}

\paragraph{Overview of the strategy} Our synthetic HTML code generation process involves two key steps for maximizing diversity and quality. First, we employ a smaller language model to generate a variety of website themes and designs. These creative outputs serve as the foundation for the next stage, where they are fed into the prompts of a larger language model, mostly trained on code data. This LLM then generates the final HTML code, ensuring that our dataset encompasses a wide range of styles while generating high-quality codes.

\paragraph{Generating diverse website concepts} We employ Mistral-7B-Instruct \citep{jiang2023mistral} to generate several million unique website concepts and designs with the prompt:

\begin{tikzpicture}
  \draw node[draw=black,fill=codebackground!50,rounded corners,inner sep=2ex,text width=\textwidth] {
    Generate diverse website layout ideas for different companies, each with a unique design element. Examples include: a car company site with a left column, a webpage footer with a centered logo. Explore variations in colors, positions, and company fields. Don't give any explanations or recognition that you have understood the request, just give the list of 10 ideas, with a line break between each.
    };
\end{tikzpicture}

Here are 5 examples of generated concepts:

\begin{tikzpicture}
  \draw node[draw=black,fill=codebackground!50,rounded corners,inner sep=2ex,text width=\textwidth] {
    1) Fashion Brand: A visually stunning layout with a full-width, rotating image carousel showcasing their latest collections, a bold, center-aligned logo, and a bottom navigation menu. The color palette is inspired by the latest fashion trends.
    
    2) Restaurant Chain: A design with a mouth-watering header image of a popular dish, a top navigation menu with pages for location, menu, and order online. The footer displays social media icons, a newsletter sign-up form, and contact information.
    
    3) Consulting Firm: A clean, professional design with a full-width image and a tagline, a left column for the main navigation menu, and a right column for featuring key staff members and their service offerings. A muted color palette evokes trust and expertise.
    
    4) Real Estate Agency: A user-friendly design with a header featuring a property image and search functionality, a top navigation menu with links to the various property listings. The page includes a grid layout for listing images, and a footer with contact information and social media icons.
    
    5) Education Platform: A design with a wide, hero image, a centered logo, and a top navigation menu featuring links for courses, pricing, and about the company. The site includes a feature section where students can share testimonials and success stories. A warm, approachable color palette is used to inspire confidence and engagement.
    };
\end{tikzpicture}

\paragraph{Opting for Tailwind CSS over traditional CSS} Generating visually diverse and appealing designs requires more than just pure HTML. However, to simplify the learning process of VLMs, employing standalone code is preferable to managing separate files. In this context, Tailwind CSS emerges as an ideal solution. This utility-first framework allows creating unique designs by providing a wide array of utility classes, enables direct styling within the HTML document, and eliminates the need for external style files. Tailwind CSS offers an extensive array of predefined classes that mirror various CSS properties. By integrating these utility classes into HTML elements, we can efficiently style web pages, resulting in concise code that is easier for VLMs to learn from.

\paragraph{Using a code specialized LLM to generate the HTML codes} To generate the final HTML codes, we leverage Deepseek-Coder-33b-instruct \citep{guo2024deepseekcoder}, a state-of-the-art language model mostly trained on code data and fine-tuned to follow instruction. We use the prompt:

\begin{tikzpicture}
  \draw node[draw=black,fill=codebackground!50,rounded corners,inner sep=2ex,text width=\textwidth] {
    Code a complete website with a good design in HTML and Tailwind CSS about this: \{concept\}\newline
    Write the code inside a tag <body>.\newline
    Write real and long sentences about the business. NEVER USE sentences starting with Lorem ipsum, NEVER.\newline
    You don't have to include images, but if you do, use only this source "https://source.unsplash.com/random/WxH/?keyword", by replacing `W` and `H` in the URL by the desired width and height, and `?keyword` by a keyword describing the picture, for example "https://source.unsplash.com/random/300x200/?gym" for an image about gym of size 300x200, or "https://source.unsplash.com/random/100x200/?cake" for an image of a cake of size 100x200.
    };
\end{tikzpicture}

An initial challenge was the text-only nature of our outputs, contrasting with the real websites containing many images. The task of integrating images into an HTML code seems hard, especially when trying to look for images related to the context of the web page. However, we discovered an effective solution through photo stocks like \texttt{https://source.unsplash.com/}, which offers the capability to dynamically generate images based on keywords, thus providing images of any size and relevant to any specified topics.

After a filtering step in which we discard web pages with insufficient text, generic content or images not aligning with the website's topic, we finally ended up with 2 million web pages.

\paragraph{Screenshot capture process} We use Playwright\footnote{\url{https://github.com/microsoft/playwright}} to visualize and capture the output of our generated HTML codes. We ensure that screenshots encompass the entire web page, regardless of its length. As a result, our dataset features screenshots in a wide range of resolutions. This diversity in image size and format is useful for enhancing the robustness of our model.

\paragraph{Visualization of WebSight examples} Five examples present in WebSight are shown in Figure \ref{fig:ex_wesight}.

\section{Fine-tuning a foundation vision-language model on WebSight}

\paragraph{Model prerequisites for webpage conversion} For a model to accurately convert webpage screenshots into HTML code, it necessitates several capabilities. These include advanced OCR to transcribe text from images, spatial understanding to arrange elements on the page, and object recognition abilities to replicate images similar to those in the input with the strategy explained above.

We use our forthcoming foundation VLM as the base model. It is built upon Mistral-7B \citep{jiang2023mistral} and SigLIP-SO400M \citep{zhai2023sigmoid}, and is using the Patch n’ Pack strategy \citep{dehghani2023patch} to preserve the original aspect ratio of the input images, with a resolution of up to 980 pixels for each side.

This base model was trained mostly on OBELICS \citep{laurencon2023obelics}, synthetic captions of image/text pairs datasets, and a combination of OCR datasets \citep{biten2022ocridl}. 

Further insights into the model’s architecture and its training process will be detailed upon its release.

\paragraph{Fine-tuning on WebSight}

For the fine-tuning, instead of unfreezing all the weights, which requires lowering significantly the learning rate for a stable training, we use the parameter efficient DoRA method \citep{liu2024dora} with a rank 64. We use the same learning rate that was chosen during the pre-training, $10^{-4}$, while seeing 2016 examples per iteration, for a total of 1100 iterations, representing a bit less than one epoch.

We find that the validation loss is not a good indicator of the trained model and in particular the quality of generated codes in real-world cases. Consequently, we perform checkpoint selection by manually inspecting generated samples rather than relying on the validation loss. Despite the validation loss continuing to decrease significantly over several epochs, it did not translate into an increased ability to generalize to websites that differ from those in the training dataset.

\section{Qualitative evaluation}

\subsection{Results on different screenshots}

\begin{figure}
    \centering
    \includegraphics[width=0.83\textwidth]{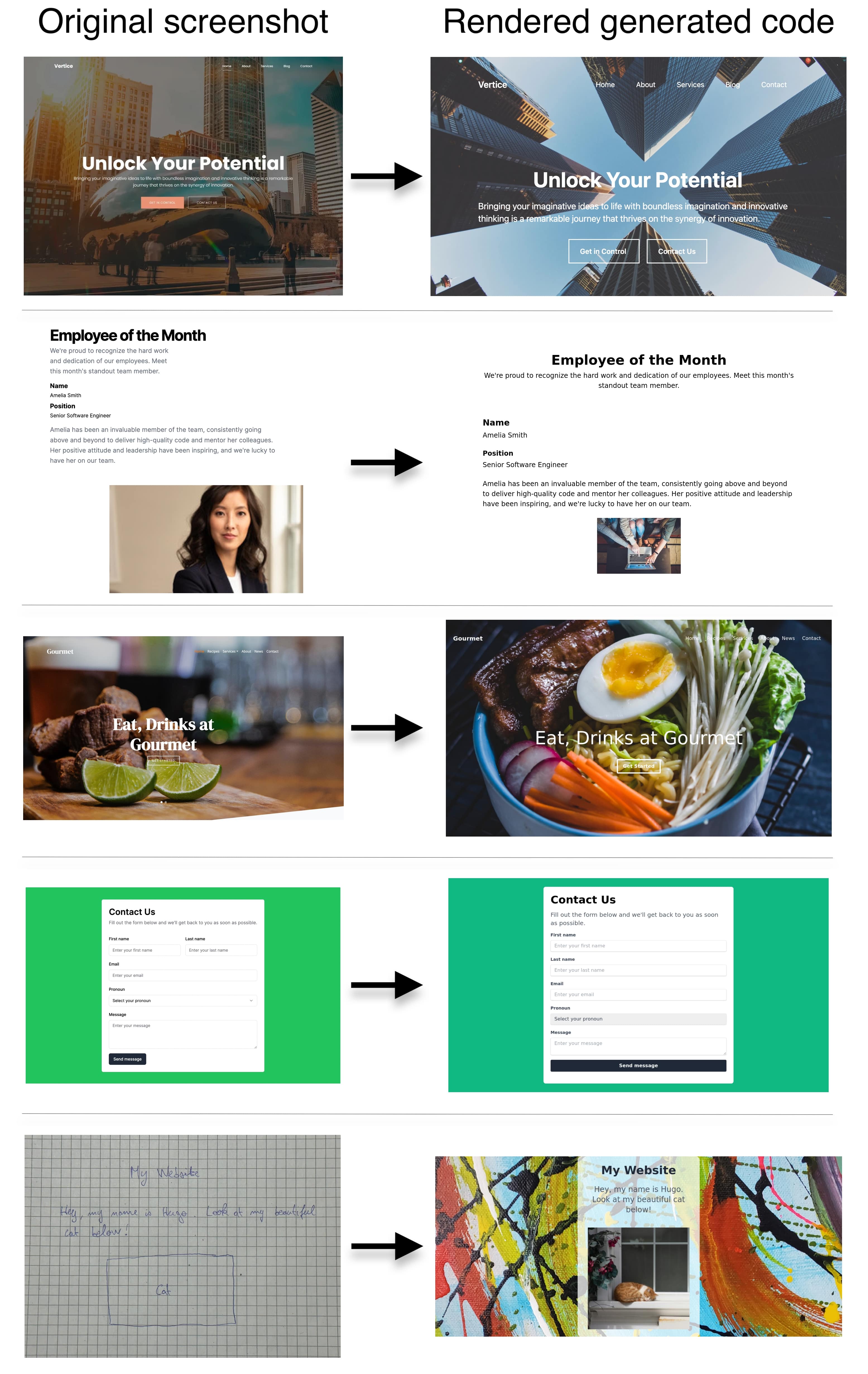}
    \caption{Comparison of an original web page (input) on the left, and the rendering of the code generated by our model, Sightseer, (output) on the right.}
    \label{fig:all_generations}
\end{figure}

Figure \ref{fig:all_generations} showcases various outputs from Sightseer when fed with simple website designs. Notably, in instances where the input contains a limited amount of text, this text tends to be accurately preserved in the output.

Remarkably, Sightseer sometimes exhibits the capability to generalize beyond its training dataset to websites that differ significantly in appearance, as evidenced by its conversion of a handwritten website sketch into functional HTML code.

\subsection{Failure cases}

\begin{figure}[h!]
    \centering
    \includegraphics[width=0.6\textwidth]{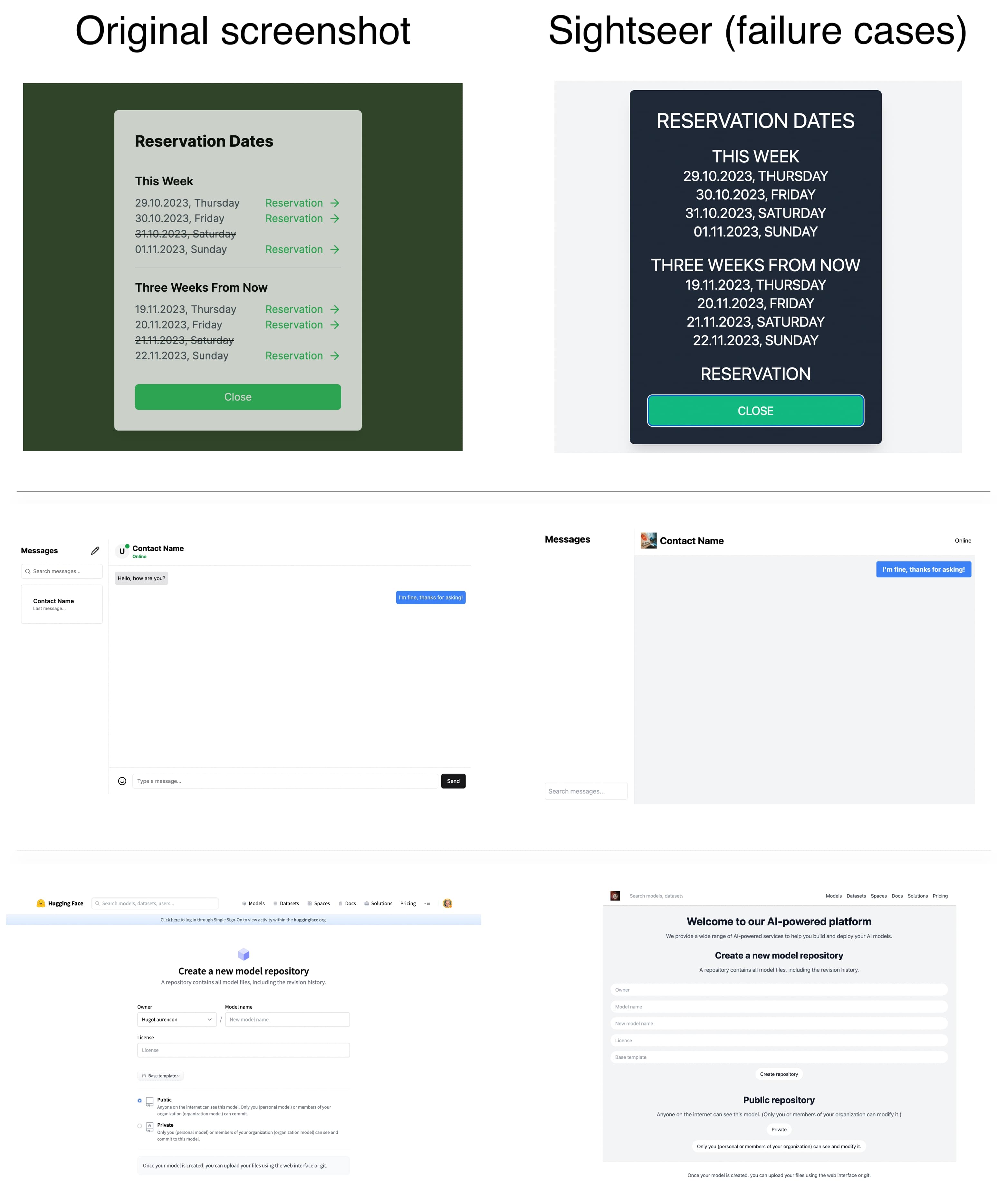}
    \caption{Examples where Sightseer-generated codes fall short in replicating the intended design.}
    \label{fig:failure_cases}
\end{figure}

In our analysis, Sightseer struggles with complex website layouts, excessive text, or designs significantly divergent from its training data.

In some instances, generated code includes elements such as images, text, or buttons that fail to appear upon rendering. This can result from issues like text colored identically to its background or incorrect syntax use, suggesting that Sightseer has not fully mastered the HTML + Tailwind CSS syntax.

While the model produces visually more attractive websites, it sometimes produces errors not observed in our initial model\footnote{\url{https://huggingface.co/HuggingFaceM4/VLM_WebSight_finetuned}} trained on WebSight-v0.1, which used traditional CSS instead of Tailwind CSS. As a more recent framework than traditional CSS, Tailwind CSS has less frequent occurrence in the pre-training data of the base LLM, and we hypothesize that the LLM has bigger challenges in fully mastering its syntax. We posit that starting with a foundational VLM pre-trained with text-only HTML + Tailwind CSS in the mixture of data could significantly enhance Sightseer's translation accuracy, and we are exploring related strategies to achieve this improvement.

\section{Conclusion}

In this work, we introduce WebSight, a large synthetic dataset of 2 million pairs of HTML codes and corresponding renderings, and Sightseer, a vision and language model with OCR ability fine-tuned on WebSight, as contributions towards automating the conversion of webpage screenshots to HTML code. By leveraging synthetic data generation and fine-tuning a high-capacity base VLM on the dataset, we demonstrate a viable path to accelerate UI development tasks and enhance no-code solutions with increasingly more powerful AI-powered tools. By open-sourcing WebSight, we aim to foster further innovation and research in this area.


\nocite{*}
\bibliographystyle{plainnat}
\bibliography{references}


\end{document}